%% file: 9303315.tex
\documentstyle[12pt]{article}
\input matex
\def\alphas{$\alpha_s(M_{^Z})$}
\def\ee{$e^+e^-$ }
\begin{document}
\begin{titlepage}
\rightline{FTUV/93-9}
\rightline{IFIC/93-5}
\rightline{February 1993}
\noindent
%\today
%\hfill Submitted to Phys. Lett. \\
\begin{center}
%\hfill prelim. version\\
{\large \bf POSSIBLE IMPLICATIONS OF A LIGHT GLUINO\\}
\vskip .4cm
{\large \bf F. de Campos}
\footnote{Bitnet CAMPOSC@EVALVX}\\
and
\vs .1cm
{\large \bf J. W. F. Valle}%%$^1$
\footnote{Bitnet VALLE@EVALUN11 - Decnet 16444::VALLE}\\
Instituto de F\'{\i}sica Corpuscular - IFIC/C.S.I.C.\\
Dept. de F\'isica Te\`orica, Universitat de Val\`encia\\
46100 Burjassot, Val\`encia, SPAIN\\
\vs .1cm
%%\vskip 0.1cm
%Theoretical Physics Division, CERN\\
%CH-1211 Geneve 23, Switzerland\\
\vskip 0.25cm
\abstract{
Light gluinos have been suggested in order to reconcile
$\alpha_s$ determinations from low energy deep inelastic
experiments with those inferred from LEP measurements.
{}From this hypothesis then one expects, in unified N=1
supergravity models, that also the "photino" will be
light. We show that this possibility is not in conflict
with recent LEP measurements and also that cosmology
does not necessarily rule it out in any convincing way.
Moreover it leads to $upper$ limits on the masses of
the other supersymmetric fermions, for example, the
lightest chargino should be lighter than about 75 GeV}
\end{center}
\noi
\vs .1cm
%\begin{flushleft}
%CERN-TH.6624/92\\
%August 1992
%\end{flushleft}
\noi
%$^1$ {\small Permanent address, Instituto de F\'{\i}sica
%Corpuscular - IFIC/C.S.I.C.}\\
%Dept. de F\'isica Te\`orica, Universitat de Val\`encia\\
%46100 Burjassot, Val\`encia, SPAIN}
\end{titlepage}
\setcounter{page}{1}
\pagestyle{plain}
%=======================================================================
There is a lot of controversy on whether or
not the existence of light gluinos, of mass
$2 \ltap m_{\tilde{g}} \ltap 6$ GeV, has been
confidently ruled out \cite{PDG92,UA1,Ant}. While this
possibility is theoretically marginal and there
is much circumstantial evidence against it,
we have been recently encouraged by the fact
that light gluinos might account for the
apparent discrepancy between the values of
the strong coupling constant $\alpha_s$
as determined from low energy deep inelastic experiments
and those inferred from high energy LEP experiments.
Indeed, the results of deep inelastic lepton-nucleon
scattering give \alphas = 0.112 $\pm$ 0.004 at the
Z mass scale, which is lower than the results of \ee
analyses of event shapes. The averaged LEP value obtained
this way is {\alphas = 0.124 $\pm$ 0.005} \cite{Bethke}.

While it is perfectly possible that the theoretical
uncertainties are underestimated in the above analyses
\cite{Altarelli} one can also envisage an alternative
solution which should not be overlooked unless one
can convincingly exclude it from experiment, namely,
the existence of an electrically neutral coloured
fermion of relatively low mass. This would slow down the
running of $\alpha_s$ between the scales accessible
at LEP and those being probed by deep inelastic
experiments. Indeed, the evolution of $\alpha_s$
between 5 and 90 GeV is described much better if
we postulate a light gluino than if we do not
\cite{JezabekKuhn}. Similar arguments were also
given in ref. \cite{Clavelli} on the basis of
quarkonia data.

If gluinos are light one also expects
the "photino" to be light. This is the
case in the most popular class of N=1 supergravity models
where there is a common supersymmetry breaking
mass parameter at the unification scale \cite{revsusy}.
In these theories most likely the lightest
of the neutralinos is the lightest supersymmetric
particle (LSP) and is mostly a "photino".
The masses and mixing angles of the charged
and neutral supersymmetric fermions, charginos
and neutralinos, are then determined by only
three independent parameters: the gluino mass,
which we may fix anywhere in the range of interest,
the ratio of the two Higgs vacuum
expectation values $\tan \beta = \frac{v_u}{v_d}$,
and the Higgsino mixing parameter $\mu$.

In this note we show that the possibility of
light gluinos in the context of supergravity
models is not in conflict with the recent
measurements of the Z decay widths,
sensitive to the possible presence of supersymmetric
Z decay channels, as well as the direct supersymmetry
searches at LEP. The relevant constraints may be
summarized as \cite{PDG92,lep}
\ben
\item
The limit on mass of the lightest of the charginos
$m_{\tilde{\chi}^{\pm}} \geq 45\mbox{ GeV}$
\item
The LEP limits on the total $Z$ width,
$\Gamma^{total}_{Z}=2.487\pm0.010\mbox{ GeV}$,
\item
The LEP limit on the invisible $Z$ decay width,
$\Gamma_{inv}=498\pm8\mbox{ MeV}$
\een
In addition to these we have also included
the LEP limit on the hadronic peak cross section
as well as the limits from $p\bar{p}$ colliders
on the ratio of W-to-Z cross sections
$0.825\leq\frac{R}{R_{SM}}\leq1.091$,
which could also be modified by the existence
of supersymmetric decay channels.

For fixed gluino mass one can determine
the region of supersymmetric parameters
allowed by the LEP experiments just in
terms of $\mu$ and $\tan\beta$
\footnote{We have not applied the constraint
$\tan\beta \geq 1$ that holds in models with
radiative electroweak breaking.}.  We have
determined what this region is for arbitrary
values of the gluino mass in the range
$ m_{\tilde{g}} \ltap 7$ GeV.
%%%%%%%%%%%%%%%%%%%%%%%%%%%%%%%%%%%%%%%%%%%%%%%%
The result is shown in figure 1. We see that
there is a finite, but nonegligible butterfly-shaped
region of allowed $\mu$ and $\tan\beta$ values.
In particular, this shows explicitly that a very
light neutralino (most likely the LSP) is
perfectly consistent with the LEP measurements
of $\Gamma_Z^{invisible}$ if the supersymmetry
breaking gaugino mass parameters are small.
The reason for this is clear: in the limit
of strictly vanishing soft-breaking gaugino
masses the LSP is a pure photino, and it is
massless, forming an unbroken supersymmetric
multiplet with the photon. Such a state is
decoupled from the Z, and therefore no limits
can be set from LEP. How about cosmology?

If the LSP is a neutralino, almost pure photino,
then the relic LSP density will be, to a very good
approximation, inversely proportional to the LSP
annihilation cross section which, in turn, is roughly
proportional to $m_{\tilde{\gamma}}^2/m_{\tilde{e}}^4$.
Requiring it not to be too large one obtains a lower
bound on the LSP mass, for this case:
\beq
m_{\tilde{\gamma}} \gsim 1 \rm{GeV} \times
({\frac{m_{\tilde{e}}}{45 \mbox{GeV}}})^2
\label{45}
\eeq
which, from the LEP bound on selectrons, leads to
$m_{\tilde{\gamma}} \gsim 1$ GeV. For such masses
the annihilation of relic photinos is sufficient
to dilute their number density to an acceptable level.

What if $m_{\tilde{\gamma}} \lsim 1$ GeV? In this case
one would have to rely on some photino decay mechanism
in order to avoid a conflict with the standard cosmological
picture. If R parity \cite{RP} is conserved, the photino will be
absolutely stable. Thus the simplest way out is to allow
for a small amount of R parity violation. Such possibility
is definitely allowed by experiment and there are several
extensions of the minimal supersymmetric standard model
where R parity violation can occurs \cite{RPP,MASI,NPBRP}.

If the photino is very light and decays with a
substantial branching ratio to photons or charged
particles, there may be strong astrophysical and
cosmological restrictions on its lifetime and mass,
following from nucleosynthesis considerations and
the isotropy of the relic photon background.
It is, however, certainly possible to obey these
restrictions in realistic models with explicitly
broken R parity. Moreover, these constraints can be
completely avoided if the low-lass photino decays
invisibly, as expected in the case of the simplest
\21 models where R parity is broken just
spontaneously \cite{MASI}. In the latter case
the photino decays mostly by majoron emission,
$\tilde{\chi}^0 \ra \nu + majoron$, and no
important constraints can be placed from
cosmology. On the other hand, its laboratory
missing energy signatures may be indistinguishable
from those it has in the minimal supersymmetric
standard model, to the extent that the invisible
decay is dominant.

We conclude that cosmological arguments can not
convincingly rule out the existence of a light photino.

We now note that in this class of supergravity models
it is possible, although marginally, to induce small
gluino and photino masses just as a result of radiative
corrections \cite{Masi}. Gluino masses around 2 GeV
may arise this way, mostly from a top-stop loop. In this
estimate the constraint on the value of $\mu$ that follows
from figure 1, $\mu \lsim 120$ GeV, plays an important role.
The required large values of $m_{top}$ are, however, not
the ones which are favored by precision data \cite{lep,B259}.
For the case of the photino, whose mass also
receives electroweak corrections \cite{Masi},
masses close to 1 GeV are possible, for
acceptable values of $m_{top}$ and $\mu$.
Higher values for gluino and photino masses
would require the existence of a primordial
nonzero supersymmetry breaking mass parameter
at the unification scale.

An interesting point to mention is that the expected
mass spectrum in our class of light-gluino supergravity
models is characterized by relatively light supersymmetric
fermions. This follows from the allowed region of $\mu$
and $\beta$ values shown in figure 1. Indeed, for such
allowed parameter values it follows that the charginos
and neutralinos should be accessible at future accelerators.
For example, figure 2 shows the region of allowed chargino
masses. Clearly, one sees that the lightest chargino
should be lighter than about 75 GeV. A similar upper
bound also applies to the next-to-lightest of the neutralinos.
Moreover, if the relic photino population disappears only due to
selectron-mediated annihilations, also selectrons would
very close to the present limit, from \eq{45}. A quick
inspection at the renormalization group equations then
shows that the other sfermions would also be light in
this case. Although squarks are somewhat heavier, due
to colour, they too should lie in the region of sensitivity
of hadron collider experiments. One way to make the model
"safer" from being experimentally disproved would be to
allow for some R parity violation, so that the photino
can decay, as described above. This would relax the
limits on the sfermions masses, allowing them to
be heavier. However, the implied upper limits on
the chargino and neutralino masses would still hold.

In conclusion we would like to stress that
neither the existing data from LEP nor
cosmological considerations preclude the
possibility that a light photino exists,
as would be expected in N=1 supergravity
models where gluinos are sufficiently light
as to play a significant role in the running
of $\alpha_s$ between 5 and 90 GeV. The resolution of the
controversy on whether or not the existence
of light gluinos can be confidently ruled out
must rest upon the results of hadron colliders
and depend on the gluino lifetime and on the
details of the strongly interacting supersymmetric
spectrum. As suggested in \cite{JezabekKuhn},
future searches for evidence of a light gluino
in 4-jet \ee or 3-jet ep events should be pursued
by the experiments at LEP and HERA. Here we have
stressed the important complementary role played
by future searches for the other electroweakly
interacting supersymmetric fermions. For example,
we have showed that the lightest chargino should
be lighter than about 75 GeV, with a similar
upper limit applying also to the neutralino
immediately heavier than the photino.
\vskip 1cm
%\vfill
\noi
{\bf Acknowledgements}\\
We thank M. Drees, G. Farrar and A. Masiero for stimulating
discussions and M. Gonzalez-Garcia for help with the
programing. The work of F. de Campos was supported by
CNPq (Brazil).

\newpage
\noi
{\bf Figure Captions}\\
{\bf Fig 1}:\\
Regions of allowed $\mu$ and $\tan\beta$ values
in unified N=1 supergravity models with light gluinos.\\
{\bf Fig 2}:\\
Region of allowed chargino masses
in light-gluino supergravity models.

\end{document}

%% file: matex.tex
\newcommand{\noi}{\noindent}
\newcommand{\bc}{\begin{center}}
\newcommand{\ec}{\end{center}}

\def\ifmath#1{\relax\ifmmode #1\else $#1$\fi}
\def\3quarter{{\textstyle{3 \over 4}}}

\def\vs{\vskip}

\def\ra{\rightarrow}

\def\lf{\leaders\hbox to 1em{\hss.\hss}\hfill}

\def\21{$SU(2) \ot U(1)$}
\def\etal{\hbox{\it et al., }}
\def\eq#1{{eq. (\ref{#1})}}

\def\ltap{\raisebox{-.4ex}{\rlap{$\sim$}} \raisebox{.4ex}{$<$}}

\def\lsim{\raise0.3ex\hbox{$\;<$\kern-0.75em\raise-1.1ex\hbox{$\sim\;$}}}
\def\gsim{\raise0.3ex\hbox{$\;>$\kern-0.75em\raise-1.1ex\hbox{$\sim\;$}}}
\def\bel{\begin{letter}}
\def\eel{\end{letter}}
\def\beq{\begin{equation}}
\def\eeq{\end{equation}}
\def\bef{\begin{figure}}
\def\eef{\end{figure}}
\def\bet{\begin{table}}
\def\eet{\end{table}}
\def\bea{\begin{eqnarray}}
\def\ba{\begin{array}}
\def\ea{\end{array}}
\def\bi{\begin{itemize}}
\def\ei{\end{itemize}}
\def\ben{\begin{enumerate}}
\def\een{\end{enumerate}}
\def\ra{\rightarrow}
\def\ot{\otimes}
\def\eea{\end{eqnarray}}

\def\np#1#2#3{           {\it Nucl. Phys. }{\bf #1} (19#2) #3}
\def\pl#1#2#3{           {\it Phys. Lett. }{\bf #1} (19#2) #3}
\def\pr#1#2#3{           {\it Phys. Rev. }{\bf #1} (19#2) #3}
\def\prep#1#2#3{         {\it Phys. Rep. }{\bf #1} (19#2) #3}

\def\n.c.#1#2#3{         {\it Nuovo Cim. }{\bf #1} (19#2) #3}
\def\r.n.c.#1#2#3{       {\it Riv. del Nuovo Cim. }{\bf #1} (19#2) #3}

\relax